\documentclass[letter]{ptptex}

\usepackage{graphicx}
\usepackage{wrapft}
\usepackage{color}

\title{
Energy dependence of $\bar{K}N$ interactions
and resonance pole of strange dibaryons
}

\author{
Yoichi \textsc{Ikeda}$^{1,2}$, 
Hiroyuki \textsc{Kamano}$^3$, 
Toru \textsc{Sato}$^4$
}

\inst{
$^1$Department of Physics, University of Tokyo, Tokyo 113-0033, Japan\\
$^2$Nishina Center for Accelerator-Based Science, Institute for Physical
and Chemical Research (RIKEN), Saitama 351-0198, Japan\\
$^3$Excited Baryon Analysis Center (EBAC), 
Thomas Jefferson National Accelerator Facility,
Newport News, Virginia 23606, USA\\
$^4$Department of Physics, Osaka University, Osaka 560-0043, Japan
}

\abst{
We study the resonance energy of the strange dibaryons
using two models with the energy-independent 
and energy-dependent potentials 
for the s-wave $\bar{K}N$ interaction,
both of which are derived
by certain reductions from the leading order term of the effective
chiral Lagrangian.
These potential models produce rather different 
off-shell behaviors of the two-body 
$\bar{K}N$ - $\pi\Sigma$ amplitudes in $I=0$ channel,
i.e.,
the model with energy-independent (energy-dependent) potential predicts 
one (two) resonance pole in the $\Lambda(1405)$ region,
while they describe the available data equally well.
We find that the energy-independent potential model
predicts one resonance pole of the strange dibaryons,
whereas the energy-dependent potential model
predicts two resonance poles:
one is the shallow quasi-bound state of the $\bar{K}NN$,
and another is the resonance of the $\pi Y N$ with
large width.
An investigation of the binding energy of the strange dibaryons will make a significant contribution
to clarify resonance structure of s-wave $\bar{K}N$ -$\pi\Sigma$
around the $\Lambda(1405)$ region.
}

\begin{document}

\maketitle

Since the deeply bound kaonic nuclear states were
predicted\cite{Aka1},
few-nucleon systems with a kaon 
have attracted increasing interest.
Among the deeply bound kaonic states, the
resonances in the $\bar{K}NN$ - $\pi YN(Y=\Sigma,\Lambda)$ system
(strange dibaryons)
are actively investigated recently,
since the three-hadron dynamics
can be theoretically handled by
the well-established Faddeev equations.
Using the coupled-channel Faddeev equations, 
two of the authors\cite{IS07,IS08} and 
Shevchenko \textit{et al}.\cite{She}
studied the possible existence and the energy of the strange 
dibaryons.
It has been found in these studies that
off-shell behaviors of the two-body 
$\bar{K}N$ - $\pi\Sigma$ amplitudes in $I=0$ channel,
which involve the $\Lambda(1405)$ resonance,
strongly affect the determination of 
the energy of the strange dibaryons.

In Refs.~\citen{IS07} and~\citen{IS08},
the energy of the strange dibaryons is evaluated by
employing the following energy-independent potential 
for the $\bar{K}N$ - $\pi Y$ subsystem
(we refer to this potential model as `E-indep' throughout this letter):
\begin{equation}
V^{\textrm{E-indep}}_{ij}(p_i,p_j)= -C_{ij}(m_i+m_j).
\label{E-indep-pot}
\end{equation}
Here indices $i,\ j$ represent the meson-baryon channel in isospin basis;
$m_i$ ($p_i$) is the meson mass (the relative momentum in 
the center of mass frame) of the channel $i$;
$C_{ij} =\lambda_{ij}/
\left(
32 \pi^2 F_{\pi}^2
\sqrt{\omega_i(p_i) \omega_j(p_j)}
\right)$
with $\omega_i(p_i)=\sqrt{m_i^2+p_i^2}$ and $F_{\pi}=92.4$ MeV.
The coefficient $\lambda_{ij}$ is determined by the 
flavor SU(3) structure constant, 
e.g., $\lambda_{\bar{K}N,\bar{K}N}=6$,
$\lambda_{\pi\Sigma,\pi\Sigma}=8$ and
$\lambda_{\bar{K}N,\pi\Sigma}=-\sqrt{6}$.
The E-indep model predicts
one resonance pole corresponding to the $\Lambda(1405)$, 
which is the quasi-bound state of the $\bar{K}N$ state,
and a virtual state
of the $\pi \Sigma$ state (see below).
The resulting strange dibaryon
is found to be
a deeply quasi-bound $\bar{K}NN$ state 
with the binding energy $B\sim 40$ - $60$ MeV.

The energy-independent potential~(\ref{E-indep-pot}) 
is derived by a reduction\cite{IS07} from 
the so-called Weinberg-Tomozawa term,~\cite{W-T}
the leading order term of the chiral effective Lagrangian:
\begin{equation}
L_{\textrm{int}} = \frac{i}{8F_\pi^2}tr(\bar{\psi}_B \gamma^\mu
                                 [[\phi,\partial_\mu \phi],\psi_B]),
\label{eq-wt}
\end{equation}
with $\psi_B$ ($\phi$) being
the octet baryon field (the octet pseudoscalar meson field).
It is also possible to
construct a model of energy-dependent potential from Eq.~(\ref{eq-wt})
following the `on-shell factorization' used in Ref.~\citen{OR98}.
The potential for the s-wave scattering is given by
\begin{eqnarray}
V^{\textrm{E-dep}}_{ij}(p_i,p_j;E)=
-C_{ij}(2E-M_i-M_j),
\label{E-dep-pot}
\end{eqnarray}
where $E$ is the total scattering energy and $M_i$ is the baryon mass
of the channel $i$.
The potential~(\ref{E-dep-pot}) depends explicitly
on the meson-baryon scattering energy.
(We refer to this potential model as `E-dep'.)
The factor $m_i+m_j$ in Eq.~(\ref{E-indep-pot}) corresponds to 
the energy dependent factor $2E - M_i - M_j$ in Eq.~(\ref{E-dep-pot}).
A common feature between the potentials of E-dep 
and the chiral unitary approach developed in
Refs.~\citen{OR98,Oller,Jido,Bor}
is that the $\bar{K}N$ interaction depends explicitly on the
scattering energy.
The chiral unitary approach predicts
two resonance poles of the $\Lambda(1405)$\cite{Jido}.

As shown below,
the E-dep model indeed predicts two resonance poles in the $\Lambda(1405)$
region and has a very different
off-shell behavior of the $\bar K N$ - $\pi\Sigma$ amplitudes 
from the E-indep model, whereas both models reproduce 
the available data of the $\bar K N$ - $\pi\Sigma$ reactions
equally well.
Our purpose of this letter is to examine
how this different nature of the $\Lambda(1405)$
in the two potential models emerges 
in the energy of the strange dibaryon
within the framework of the Faddeev equations.


\vspace{1cm}
Before examining the resonance poles
in the $\bar{K}NN$ - $\pi Y N$ system,
we first describe the two reaction models, E-indep and E-dep, for the
$I=0$ $\bar{K}N$ - $\pi \Sigma$ subsystem.
The amplitudes of the E-indep (E-dep) model is obtained by solving
the Lippmann-Schwinger equations with
$V^{\textrm{E-indep}}_{ij}(p_i,p_j)$
$\left(V^{\textrm{E-dep}}_{ij}(p_i,p_j,E)\right)$ as a driving term.
In the actual calculations,
we regularize the loop integrals of the Lippmann-Schwinger 
equation with the dipole form factors 
$g_i(\vec p_i)=\Lambda_i^4/(p_i^2 + \Lambda_i^2)^2$, where
the $\Lambda_i$ is the cutoff parameter in the channel $i$.

It is noted that we will take the non-relativistic kinematics
throughout this work.
This is because the energy-dependent potential such as Eq.~(\ref{E-dep-pot})
causes a problem 
in the three-body calculations with the relativistic kinematics:
Two-body scattering energy $W_2$,
which is defined by 
$W_2 = \sqrt{ (W - E_{\textrm{sp}})^2 - p_{\textrm{sp}}^2 }$ with
the total scattering energy $W$ and
the spectator energy $E_{\textrm{sp}}$ and momentum $p_{\text{sp}}$,
becomes pure imaginary for large spectator
momenta.
With the non-relativistic kinematics
such a difficulty can be avoided.

\begin{figure}[t]
\begin{center}
\includegraphics[clip,width=0.45\textwidth]{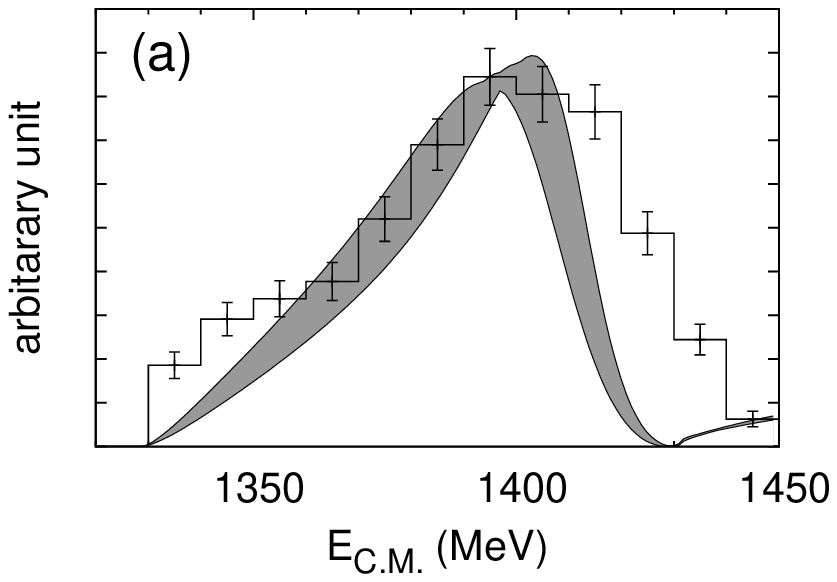}
~
\includegraphics[clip,width=0.45\textwidth]{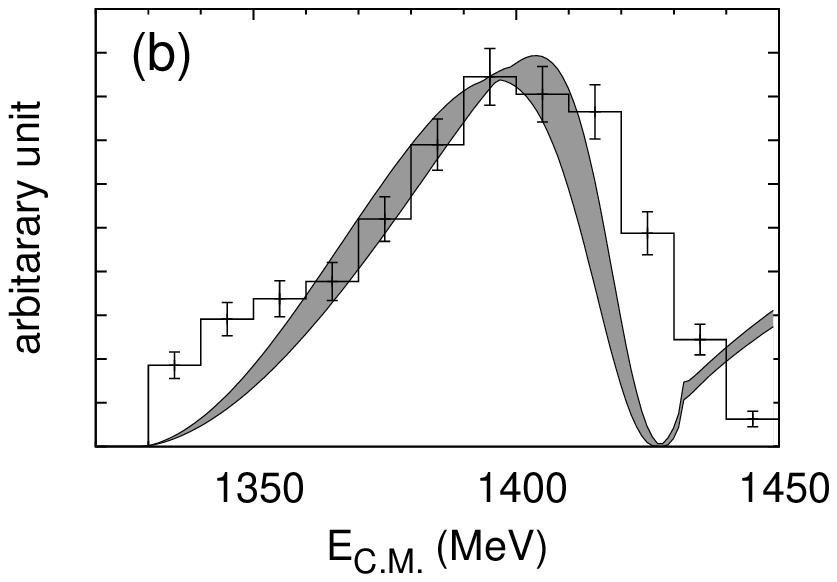}
\caption{
The invariant mass distributions of 
the $I=0$ $\pi \Sigma$ for (a) E-indep and (b) E-dep models.
}
\label{fig1}
\end{center}
\end{figure}

We determine the 
cutoff parameters 
by fitting to the shape of 
the $\pi\Sigma$ invariant mass distribution in the
$K^-p \rightarrow \pi \pi \pi \Sigma$ reaction\cite{hem}
obtained with the assumption, 
$dN/dE \varpropto |t_{\pi\Sigma,\pi\Sigma}|^2 p_{\pi\Sigma}$
($p_{\pi\Sigma}$ denotes the momentum in the $\pi\Sigma$ center of mass frame).
The results of the fit are shown in Fig.~\ref{fig1} and
the ranges of cutoff parameters are listed in Table~\ref{table1}.
It is noted that we only vary the parameters
within a range of the uncertainty of the data.
\begin{table}[b]
\begin{center}
\caption{
The cutoff parameters of the E-indep and E-dep models.
In the forth column, the poles in the $\bar{K}N$ - $\pi \Sigma$ amplitudes
on the $\bar{K}N$ physical and the $\pi \Sigma$ unphysical energy sheet
are also shown.
}
\begin{tabular}{cccc}
\hline \hline
 & $\Lambda_{\bar{K}N}$ (MeV) & $\Lambda_{\pi \Sigma}$ (MeV) &
 Poles (MeV)\\
\hline
E-indep & 975 - 1000 & 675 - 725  & 
$(1405$ - $1411)-i(14$ - $20)$, $(1296$ - $1306)-i0$\\
E-dep   & 975 - 1000 & 675 - 725  & 
$(1417$ - $1423)-i(16$ - $21)$, $(1335$ - $1341)-i(65$ - $79)$\\
\hline
\end{tabular}
\label{table1}
\end{center}
\end{table}

\begin{figure}[b]
\begin{center}
\includegraphics[width=0.45\textwidth]{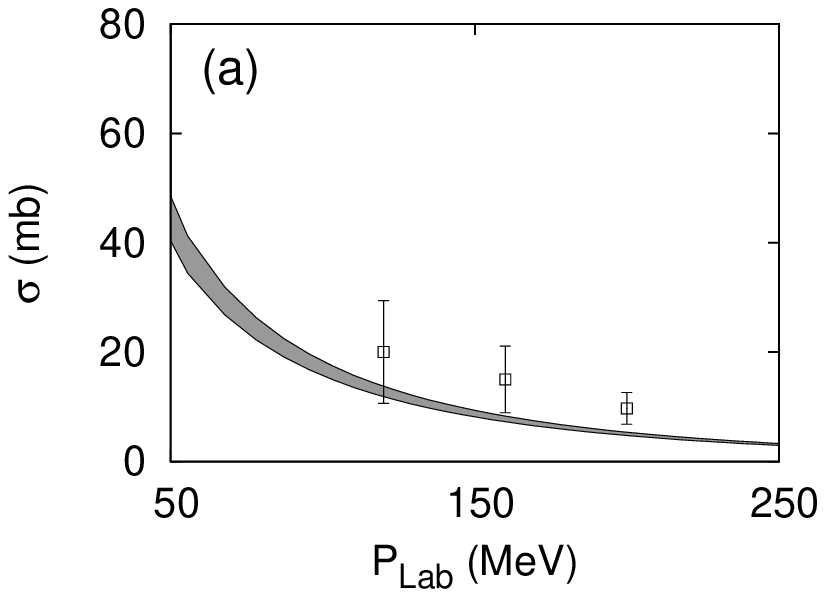}
~
\includegraphics[width=0.45\textwidth]{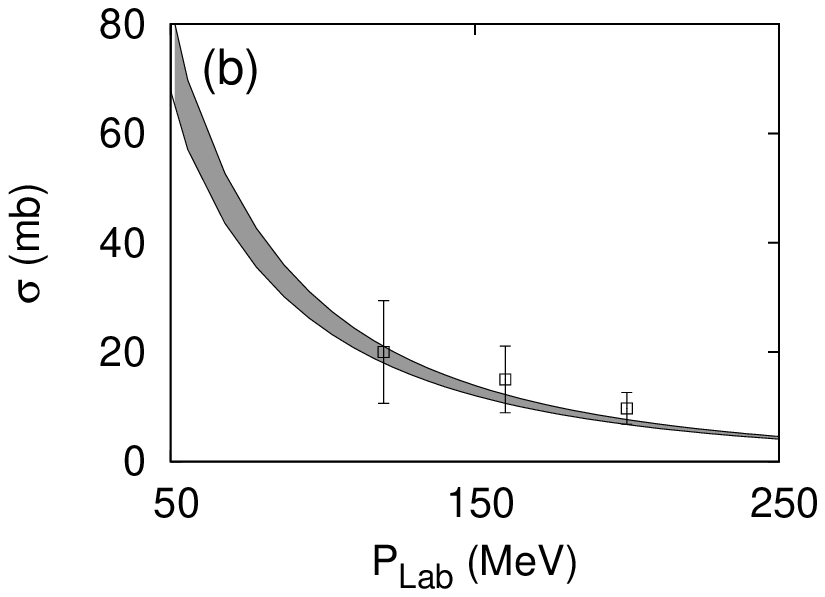}
\caption{
The total cross section of the $K^-p \rightarrow \pi^0\Sigma^0$ reaction of
(a) E-indep and
(b) E-dep models
}
\label{fig2}
\end{center}
\end{figure}
\begin{figure}
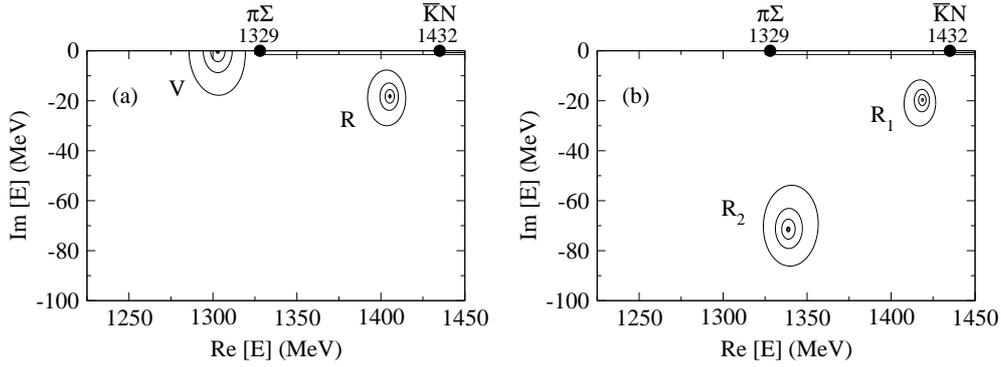

\begin{center}
\includegraphics[clip,width=0.45\textwidth]{fig3a.eps}
~
\includegraphics[clip,width=0.45\textwidth]{fig3b.eps}
\caption{
Contour plots of the absolute values of t-matrix
in $I=0$ $\bar{K}N$ - $\pi \Sigma$ system
for (a) E-indep and (b) E-dep models
with cutoff parameters 
$(\Lambda_{\bar{K}N},\ \Lambda_{\pi \Sigma}) =
(1000,\ 700)$ MeV.
The $\pi \Sigma$ and $\bar{K} N$ threshold energy
are shown together.
}
\label{fig3}
\end{center}
\end{figure}
With the parameters obtained by the fit, we
also calculate
the total cross sections of $K^- p \rightarrow \pi^0 \Sigma^0$, 
to which only the $I=0$ $\bar{K}N-\pi\Sigma$ system contributes.
The resulting $K^- p \rightarrow \pi^0 \Sigma^0$ total cross sections
are shown in Fig.~\ref{fig2}.

In Fig.~\ref{fig3} and Table~\ref{table1},
we present the poles in the two-body amplitudes
obtained from the E-indep and E-dep models.
In the E-indep model (Fig.~\ref{fig3} (a)), 
we find one resonance pole, which corresponds 
to the $\Lambda(1405)$, 
in $(1405$ - $1411)-i(14$ - $20)$ MeV ($R$)
on the $\bar{K}N$ physical
and the $\pi\Sigma$ unphysical energy sheet.
The pole $R$ becomes the $\bar{K}N$ bound state
if we turn off the $\bar{K}N$ - $\pi\Sigma$ transition potential.
In addition, we find a virtual state in  $(1296$ - $1306)$ MeV ($V$)
on the same energy sheet.

On the other hand, 
the E-dep model has two resonance poles
in the $\Lambda(1405)$ region
in $(1417$ - $1423)-i(16$ - $21)$ MeV ($R_1$) and 
$(1335$ -$ 1341)-i(65$ - $79)$ MeV ($R_2$)
as shown in Fig.~\ref{fig3} (b) and Table~\ref{table1}.
If the transition potentials between $\bar{K} N$ - $\pi\Sigma$ channels
are turned off, the poles $R_1$ become 
the $\bar{K}N$ bound states 
and $R_2$ become the resonances in the $\pi\Sigma$ channel.
The similar result has been reported in Ref.~\citen{HW}.

We observe that 
the poles which strongly couple to 
the $\bar{K}N$ channel, $R$ and $R_1$, are rather stable against
rather large variation of the parameters
as far as they describe equally well the available data of
the $\bar KN$ - $\pi\Sigma$ scattering at low energies.
On the other hand,
the pole positions of $V$ and $R_2$ 
which strongly couple to the $\pi \Sigma$ channel
depend on reaction models.
It is worthwhile to mention that 
the $\pi \Sigma$ scattering length may provide
a useful constraint on the determination of 
the poles that strongly couple to the $\pi \Sigma$ channel.\cite{JSIHO}


\vspace{1cm}
Now let us discuss the resonance poles of the strange dibaryons
in the E-indep and E-dep models. 
We shall show that
the difference in the off-shell behavior of $\bar{K}N$ - $\pi\Sigma$ amplitude
strongly affects the resonance energy of the strange dibaryons
in spin-parity $J^{\pi}=0^-$ and isospin $I=1/2$ channel.

In the three-body calculation,
we start with
the coupled-channel Alt-Grassberger-Sandhas (AGS) equations~\cite{ags}
for the $\bar{K}NN$ - $\pi YN$ system.
The equation is given by
\begin{eqnarray}
X_{ij}(p_i,p_j;W)=C^1_{ij}Z_{ij}(p_i,p_j;W)+
\sum_{l,m} \int dq q^2 C^2_{il} Z_{il}(p_i,q;W)
\tau_{lm}(W) X_{mj}(q,p_j;W),\nonumber \\
\label{cc-AGS}
\end{eqnarray}
where $i,j$ are specified by the Fock space of
the three particles and the quantum number of the interacting pair,
and $W$ is the three-body scattering energy.
$X$, $Z$ and $\tau$ denote the three-body amplitudes,
particle-exchange potentials and isobar propagators, respectively.
The coefficients $C^{1,2}$ are the spin-isospin re-coupling
coefficients given in Ref.~\citen{IS07}.
We here define the two-body scattering energy $W_2$ 
in the three-body system as
\begin{eqnarray}
W_2 = W - m_i - \frac{\vec p_i^2}{2 \eta_i},
\end{eqnarray}
where $m_i$, $\vec p_i$ and $\eta_i$ are the mass, momentum and
reduced mass of the spectator particle $i$, respectively.
As mentioned above, this non-relativistic 
energy helps to avoid the problem
of the energy-dependent potentials.
The cutoff parameters of the meson-baryon interaction 
for the E-indep and E-dep model are summarized in Table~\ref{table2-2}.
\begin{table}[t]
\begin{center}
\caption{The cutoff parameters for the 
$I=0$ and $I=1$ $\bar{K}N-\pi Y$ scatterings,
and the $I=1/2$ and $I=3/2$ $\pi N$ scatterings.
The values are given in MeV.
}
\begin{tabular}{cccccccc}
\hline \hline
& $\Lambda^{I=0}_{\bar{K}N}$ & $\Lambda^{I=0}_{\pi \Sigma}$ & 
 $\Lambda^{I=1}_{\bar{K}N}$ & $\Lambda^{I=1}_{\pi \Sigma}$ & 
 $\Lambda^{I=1}_{\pi \Lambda}$  & 
$\Lambda^{I=1/2}_{\pi N}$ & $\Lambda^{I=3/2}_{\pi N}$ \\
\hline
E-dep   &975 - 1000 & 675 - 725 & 920 & 960 & 640 & 400 & 400 \\
E-indep &975 - 1000 & 675 - 725 & 725 & 725 & 725 & 400 & 400 \\
\hline
\end{tabular}
\label{table2-2}
\end{center}
\end{table}
We take into account the Yamaguchi type of 
the baryon-baryon interactions 
for $NN$ in $^1 S_0$ channel\cite{IS07}
and $YN$ in isospin $I=1/2$ and $3/2$ channels\cite{YN}.
It is noted that we again give ranges for the cutoff parameters
of $\bar KN$ - $\pi\Sigma$ channels.
As a result, the energy of the strange dibaryons shown below
is also presented with a range.

Solving the coupled-channel AGS equations~(\ref{cc-AGS})
we find
one resonance pole for the E-indep model
on the $\bar{K}NN$ physical and $\pi YN$ unphysical sheet.
The resonance energy is
$(B, \Gamma/2)=(44$ - $58,17$ - $20)$ MeV, which is expressed
with the binding energy $B$ from the $\bar{K}NN$ threshold
and the width $\Gamma$.
In contrast,
deeper binding energy ($B \sim 45$ - $95$ MeV) of the strange dibaryon
is predicted 
with the use of the relativistic kinematics
in Ref.~\citen{IS07}.

On the other hand,
we find
two resonance poles of the strange dibaryons for the
E-dep model
on the $\bar{K}NN$ physical and $\pi YN$ unphysical sheet.
One pole
(Pole I) is found in
$(B, \Gamma/2)=(9$ - $16,17$ - $23)$ MeV, 
which
is small binding energy and relatively small width.
This implies that the strange dibaryon associated with
Pole I will be observed as a shallow bound state.
The similar result ($B \sim 20$ MeV) with
the variational calculation
was reported in Ref.~\citen{DHW}.
The other pole
(Pole II) is found at
$(B,\Gamma/2)=(67$ - $89,122$ - $160)$ MeV.
Since the imaginary part of Pole II is large,
it may be hard to identify the resonance associated
with Pole II in future experiments.
Experimental studies of the binding energy of the strange dibaryon
would give information on the subthreshold properties of the $\bar{K}N$
amplitude,
which are not well constrained from the currently available
data.

\vspace{1cm}

In summary, we have
examined how the different nature of the $\Lambda(1405)$
emerges in the energy of the strange dibaryon
within the framework of the Faddeev equations.
First we
studied the resonance poles of the $\Lambda(1405)$
and the strange dibaryon using the
models with energy-independent and energy-dependent
potentials for the s-wave $\bar KN$ - $\pi\Sigma$ interactions.
The meson-baryon interactions have been
derived form the Weinberg-Tomozawa term of the 
effective chiral Lagrangian.
Only one resonance pole corresponding to the $\Lambda(1405)$
is predicted in the model with the energy-independent potential,
whereas the energy-dependent potential model
predicts the two resonance poles.
The difference is the appearance of 
resonance pole which strongly couples to
the $\pi \Sigma$ channel.
The appearance of this resonance pole
is attributed to the energy dependence of the potentials.

We then studied the strange dibaryons.
The energy-dependent potential model predicts two 
poles of the strange dibaryons:
one is the shallow bound state and another
is the state with very large width.
The latter state, which is  
originated in the two-body $\pi \Sigma$ resonance, however,
may be difficult to be observed experimentally.
Meanwhile the energy-independent potential\cite{IS07,IS08}
predicts the deeper binding energy
than that of the energy-dependent potential model.
Therefore, experimental determination of the 
binding energy of the strange dibaryon will make a significant contribution
to clarify resonance structure of s-wave $\bar{K}N$ -$\pi\Sigma$
around $\Lambda(1405)$ region.

\vspace{1cm}
\begin{center}
{\bf Acknowledgement}
\end{center}
\vspace{1cm}

This work is supported 
by the Japan Society for the Promotion of Science,
Grant-in-Aid for Scientific Research on Innovative Areas (No. 2004:
20105001, 20105003), 
Grant-in-Aid for Scientific Research(C) 20540270,
and by 
the U.S. Department of Energy, Office of Nuclear Physics Division, 
under Contract No. DE-AC05-06OR23177
under which Jefferson Science Associates operates the Jefferson Lab.

\end{document}